# Delay Analysis for Wireless Local Area Networks with Multipacket Reception under Finite Load

Ying Jun (Angela) Zhang, Member, IEEE, Soung Chang Liew, Senior Member, IEEE, and Da Rui Chen
Student Member, IEEE
*Dept. of Information Engineering, The Chinese University of Hong Kong*
*Email: {yjzhang, soung, drchen}@ie.cuhk.edu.hk*

***Abstract -*** To date, most analysis of WLANs has been focused on their operation under saturation condition. This work is an attempt to understand the fundamental performance of WLANs under unsaturated condition. In particular, we are interested in the delay performance when collisions of packets are resolved by an exponential backoff mechanism. Using a multiple-vacation queueing model, we derive an explicit expression for packet delay distribution, from which necessary conditions for finite mean delay and delay jitter are established. It is found that under some circumstances, mean delay and delay jitter may approach infinity even when the traffic load is way below the saturation throughput. Saturation throughput is therefore not a sound measure of WLAN capacity when the underlying applications are delay sensitive. To bridge the gap, we define safe-bounded-mean-delay (SBMD) throughput and safe-bounded-delay-jitter (SBDJ) throughput that reflect the actual network capacity users can enjoy when they require bounded mean delay and delay jitter, respectively.

The analytical model in this paper is general enough to cover both single-packet reception (SPR) and multi-packet reception (MPR) WLANs, as well as carrier-sensing and non-carrier-sensing networks. We show that the SBMD and SBDJ throughputs scale super-linearly with the MPR capability of a network. Together with our earlier work that proves super-linear throughput scaling under saturation condition, our results here complete the demonstration of MPR as a powerful capacity-enhancement technique for both delay-sensitive and delay-tolerant applications.

# I. INTRODUCTION

## A. Motivation and Summary of Contribution

Thanks to its simplicity, robustness and cost effectiveness, wireless local area networks (WLANs) are playing a major role in next-generation home networks and hot spots. Since the seminal work of Bianchi in [2], there have been extensive efforts in characterizing the performance of WLANs based on IEEE 802.11 distributed control function (DCF). The majority of existing work focuses on performance analysis when the network is operated under saturation condition, where stations are never idle and always have packets to be transmitted [2-5, 7]. By contrast, the fundamental characteristics of unsaturated WLANs are yet to be understood. In fact, this paper reveals the fact that under many circumstances, it is necessary to operate a WLAN *far below* the saturation load to avoid excessively long packet delay. Considering the increasing demand of delay-sensitive services in next-generation WLANs, it is crucial to understand the following questions: (i) what is the delay performance in unsaturated WLANs and how is packet delay affected by the exponential backoff (EB) associated with the MAC protocol; and (ii) what is the maximum throughput that can guarantee finite mean delay and delay jitter and how to maximize this throughput. This paper is an attempt to address these questions.

In conventional WLANs, collision of packets occurs when more than one station transmits at the same time, causing a waste of bandwidth. With PHY-layer advanced signal processing techniques such as multiuser detection (MUD) [13], it is possible for a receiver to receive multiple packets simultaneously without causing collisions. Our work in [10, 11, 14] shows that WLAN capacity can be improved greatly with multi-packet reception (MPR) enhancements in the physical (PHY) layer: saturation throughput scales super-linearly with the MPR capability of the channel. A natural question that arises is whether MPR is still a powerful capacity-enhancement technology in WLANs under unsaturated traffic condition.

To address the above important issues, this paper performs a detailed study for WLANs with MPR enhancement under non-saturation condition. With MPR, a station can successfully receive $M$ packets at a time, where $M$ denotes the number of packets that can be resolved simultaneously [9, 14]. In a special case where $M=1$, our system reduces to conventional WLANs, which is referred to as single-packet-reception (SPR) WLANs hereafter. The key contributions of this paper are as follows:

- We propose a multiple-vacation queueing model to derive the explicit expressions for the probability distribution (in terms of transform) of packet delay. The analytical model is sufficiently general to cover both EB-based carrier-sensing and non-carrier-sensing networks including slotted ALOHA with EB, IEEE 802.11 DCF basic access mode, and IEEE 802.11 DCF request-to-send/clear-to-send (RTS/CTS) mode.

- We establish sufficient and necessary conditions for finite mean delay and delay jitter. We find that under some circumstances, mean delay and delay jitter becomes unbounded even when traffic load is far below the saturation point; while in other cases, a close-to-saturation traffic load can be sustained without driving the first two moments of packet delay to infinity. This paper provides a theoretical explanation of the phenomena. In particular, the analysis in this paper shows that whether a traffic load will result in infinite delay highly depends on the backoff factor, MPR capability, as well as the relative durations of idle, collision, and success time slots.

- In contrast to existing definition of MAC throughput [2], we define safe-bounded-mean-delay (SBMD) throughput and safe-bounded-delay-jitter (SBDJ) throughputs as the maximum safe throughputs that can be sustained with finite mean delay and delay jitter, respectively (the qualifying term "safe" in SBMD and SBDJ will be elaborated in Section IV). Arguably, SBMD throughput and SBDJ throughput are more meaningful than saturation throughput in measuring the capacity of WLANs, as bounded delay is crucial to many WLAN applications. We show that similar to the saturation throughput studied in [10, 11, 14], SBMD and SBDJ throughputs scale super-linearly with MPR capability of the channel. That is, SBMD and SBDJ throughput divided by $M$ increases with $M$. This provides a strong incentive to deploy MPR in next-generation wireless networks, as the system throughput per unit cost increases with MPR capability of the channel.

- Most existing EB-based protocols adopt a backoff exponent $r = 2$. That is, when packets collide, the collision windows of the involved stations are doubled. We show in this paper that binary EB often does not yield the optimal SBMD and SBDJ throughput. Compared with saturation throughput, SBDM and SBDJ throughputs are more sensitive to $r$. This implies that one should be more careful in setting the right $r$ if the application is delay sensitive. Fortunately, with MPR, a large $M$ can decrease the sensitivity of throughputs to $r$. This provides another incentive to deploying MPR: the system is more robust against imprecise $r$ setting.

*B. Related Work*

Previous work on delay analysis has primarily focused on medium-access delay of head-of-line (HOL) packets under saturation condition [3-5, 7]. In particular, mean saturation delay can be easily derived from the reciprocal of saturation throughput [3, 4]. In [5], a 3-D Markov chain model is proposed to analyze the distribution of medium-access delay. However, no closed-form expressions are given therein. More recently, Sakurai and Vu proposed a stochastic model to obtain explicit expressions for the first two moments and the generating function of medium-access delay under saturation condition [7]. The authors proved that the moments of access delay are finite only when certain conditions are satisfied. In our paper, a similar methodology is adopted to analyze the service time in the proposed multiple-vacation queueing model. In particular, we find that the necessary conditions for finite moments of access delay in [7] can be extended to WLANs with unsaturated traffic load and MPR capability. This result can be immediately used to derive the necessary conditions for finite mean delay and delay jitter in MPR WLANs, which leads to the definition of SBMD and SBDJ throughputs in this paper.

Delay performance in unsaturated systems is not as well studied as in saturated systems. In [12], Tickoo and Sikdar presented a preliminary study of queueing delay in SPR WLANs. Probability generating function (PGF) of system time is derived when packet arrivals occur only at the beginning of a time slot. For cases where packets can arrive in the middle of a slot, only mean system time is derived. In contrast, our paper proposes a multiple-vacation queueing model to accurately model packet delay when packet arrivals occur at arbitrary time instants. Another attempt in characterizing the performance of unsaturated SPR WLANs was made in [8]. The authors observed that in IEEE 802.11 WLANs, maximum throughput occurs in non-saturation case rather than saturation case. Our work in this paper studies the fundamental theory behind the phenomenon. In particular, we find that the observation in [8] is not always the case. Whether unsaturated throughput can be higher than saturated throughput highly depends on the backoff factor and relative length of time slots. Furthermore, unlike what was suggested in [8], we find that any throughput higher than saturated throughput cannot be safely sustained. It is therefore suggested not to load the system higher than saturation throughput, even if a higher throughput is achievable in theory.

Previous work on delay analysis in MPR networks has focused on pure slotted ALOHA systems [9]. Apart from the underlying MAC protocol, a major difference between this work and [9] lies in the definition of MPR. In [9], average delay is characterized for a subclass of MPR channels, namely capture channels, where at most one user has a successful packet transmission when multiple packets contend for the channel at the same time.

The remainder of this paper is organized as follows. The system model is introduced in Section II. In Section III, we show that packet delay in WLANs can be analyzed using a multiple-vacation queueing model by properly defining service time. The probability distribution of packet delay is derived and the conditions under which mean delay and delay jitter are bounded are studied. In Section IV, we show that by varying backoff exponent and relative lengths of timeslots, delay-throughput curves exhibit different characteristics. Based on the observation, SBMD and SBDJ throughputs are defined. Numerical results show that the maximum SBMD and SBDJ throughputs scale super-linearly with the MPR capability of the channel. Furthermore, the sub-optimality of binary EB is illustrated. In Section V, numerical results of two example systems are given to further illustrate our analysis. Finally, the paper is concluded in Section VI.

## II. SYSTEM MODEL

### A. System Setup

We consider a network with $N$ stations, each having a queue. Packets arrive at each station according to a Poisson process at a rate of $\lambda$ packets per second. We assume that the channel has the capability to accommodate up to $M$ simultaneous packet transmissions thanks to advanced PHY-layer signal processing techniques. In other words, packets can be received correctly whenever the number of simultaneous transmissions is no larger than $M$. When more than $M$ stations contend for the channel at the same time, collision occurs and no packet can be decoded. By letting $M = 1$, the system reduces to a traditional SPR WLAN.

The transmission of stations is coordinated by an EB mechanism. The EB mechanism adaptively tunes the transmission probability of a station according to the traffic intensity of the network. It works as follows. At each packet transmission, a station sets its backoff timer by randomly choosing an integer within the range $[0, W-1]$, where $W$ denotes the size of the contention window. The backoff timer is decreased by one following each time slot. The station transmits a packet from its queue once the backoff timer reaches zero. At the first transmission attempt of a packet, $W$ is equal to $W_0$, the minimum contention window. Each time the transmission is unsuccessful, $W$ is multiplied by a backoff factor $r$. That is, the contention window size $W_i = r^i W_0$ after $i$ successive transmission failures. For simplicity, we assume there is no retry limit in our system. However, our analysis can be easily extended to the case with a retry limit.

In this paper, we consider WLANs under unsaturated condition, where the queues of stations are empty from time to time. Let $p_t$ be the probability that a backlogged station transmits in a generic (i.e., randomly chosen) time slot. Then, the probabilities that a generic time slot is an idle slot, collision slot, and success slot are given in the following equations, where $X$ is a random variable representing the number of the number of transmission attempts in one slot. In the following equations, the superscript $G$ in $P_{idle}^G$, $P_{coll}^G$ and $P_{succ}^G$ stands for "generic".

$$\begin{aligned} P_{idle}^G &= \Pr\{X = 0\} \\ &= \sum_{j=0}^{N} \binom{N}{j} \rho^j (1-\rho)^{N-j} (1-p_t)^j \\ &= (1-\rho p_t)^N = (1-\tau)^N \end{aligned} \quad (1)$$

$$\begin{aligned} P_{coll}^G &= \Pr\{X > M+1\} \\ &= \sum_{j=M+1}^{N} \sum_{k=M+1}^{j} \binom{N}{j}\binom{j}{k} \rho^j (1-\rho)^{N-j} p_t^k (1-p_t)^{j-k} \\ &= \sum_{k=M+1}^{N} \sum_{j=k}^{N} \binom{N}{j}\binom{j}{k} (\rho p_t)^k (1-\rho)^{N-j} (\rho - \rho p_t)^{j-k} \\ &= \sum_{k=M+1}^{N} \binom{N}{k} \tau^k (1-\tau)^{N-k} \end{aligned} \quad (2)$$

$$\begin{aligned} P_{succ}^G &= \Pr\{X \leq M\} \\ &= 1 - P_{idle}^G - P_{coll}^G \\ &= \sum_{k=1}^{M} \binom{N}{k} \tau^k (1-\tau)^{N-k} \end{aligned} \quad (3)$$

In the above, $\tau = \rho p_t$ denotes the average transmission probability of any station in a generic time slot. A close observation of (1)-(3) indicates that $\Pr\{X = k\}$ can be easily calculated as

$$\Pr\{X = k\} = \binom{N}{k} \tau^k (1-\tau)^{N-k}. \quad (4)$$

*B. Length of Backoff Timeslots*

In WLANs, the length of a time slot is not necessarily fixed and may vary under different contexts [2]. We refer to this variable-length slot as backoff slot hereafter. Let $T_{idle}$ denote the length of an idle time slot when nobody transmits; $T_{coll}$ denote the length of a collision time slot when more than $M$ stations contend for the channel; and $T_{succ}$ denote the length of a time slot due to successful transmission when the number of transmitting stations is anywhere from 1 to $M$. The durations of $T_{idle}$, $T_{coll}$, and $T_{succ}$ depend on the underlying WLAN configuration. For EB-based slotted ALOHA systems, the duration of all backoff slots are equal to the transmission time of a packet. That is,

$$T_{idle} = T_{coll} = T_{succ} = H + PL/\text{data rate}, \tag{5}$$

where $H$ is the transmission time of PHY header and MAC header and $PL$ denotes the payload length. In IEEE 802.11 DCF basic access mode,

$$\begin{cases} T_{idle} = \sigma \\ T_{coll} = H + PL/\text{data rate} + DIFS \\ T_{succ} = H + PL/\text{data rate} + SIFS + ACK + DIFS \end{cases} \tag{6}$$

where $\sigma$ is the time needed for a station to detect the packet transmission from any other station and is typically much smaller than $T_{coll}$ and $T_{succ}$; $ACK$ is the transmission time of an ACK packet; $\delta$ is the propagation delay; and $SIFS$ and $DIFS$ are the inter-frame space durations. Similarly, in DCF request-to-send/clear-to-send (RTS/CTS) access scheme, the slot durations are given by

$$\begin{cases} T_{idle} = \sigma \\ T_{coll} = RTS + DIFS \\ T_{succ} = RTS + CTS + H + PL/\text{data rate} + 3SIFS + ACK + DIFS \end{cases} \tag{7}$$

where $RTS$ and $CTS$ denote the transmission time of RTS and CTS packets, respectively.

*C. Throughput and Operating Point*

WLAN throughput $S$, defined as the average number of information bits successfully transmitted per second, can be calculated as

$$S = \frac{\sum_{k=1}^{M} k \Pr\{X = k\} PL}{P_{idle}^G T_{idle} + P_{coll}^G T_{coll} + P_{succ}^G T_{succ}}, \tag{8}$$

According to (8), throughput $S$ can be plotted as a function of transmission probability $\tau$, as illustrated in Fig. 1. The location and skewness of the curve depend on the relative length of time slots $T_{idle}$, $T_{coll}$, and $T_{succ}$ as well as the MPR capability $M$. The maximum possible throughput, denoted by $S^*$, occurs when the transmission probability is

$$\tau^* = \arg\max \frac{\sum_{k=1}^{M} k \binom{N}{k} \tau^k (1-\tau)^{N-k}}{P_{idle}^G T_{idle} + P_{coll}^G T_{coll} + P_{succ}^G T_{succ}}. \tag{9}$$

In practice, WLANs are seldom operated at $\tau^*$ unless the backoff factor $r$ is optimized. Prior work in [2, 4, 7, 14] has shown that under saturation condition when stations are continuously backlogged, $\tau$ is determined by backoff factor $r$. In particular, it is the root of a fixed-point system

$$\tau = \frac{2(1-rp_c)}{W_0(1-p_c)+1-rp_c}, \tag{10}$$

where $p_c$ is the probability that a station encounters collisions when it transmits, which is given by

$$p_c = 1 - \sum_{k=0}^{M-1} \binom{N-1}{k} \tau^k (1-\tau)^{N-1-k}. \tag{11}$$

In the rest of the paper, we denote the transmission probability under saturation by $\tau_s$ and the corresponding throughput by $S_s$. Depending on $r$ and $M$, $\tau_s$ can be smaller than, equal to, or larger than $\tau^*$. The case where $\tau_s < \tau^*$ is illustrated in Fig. 1.

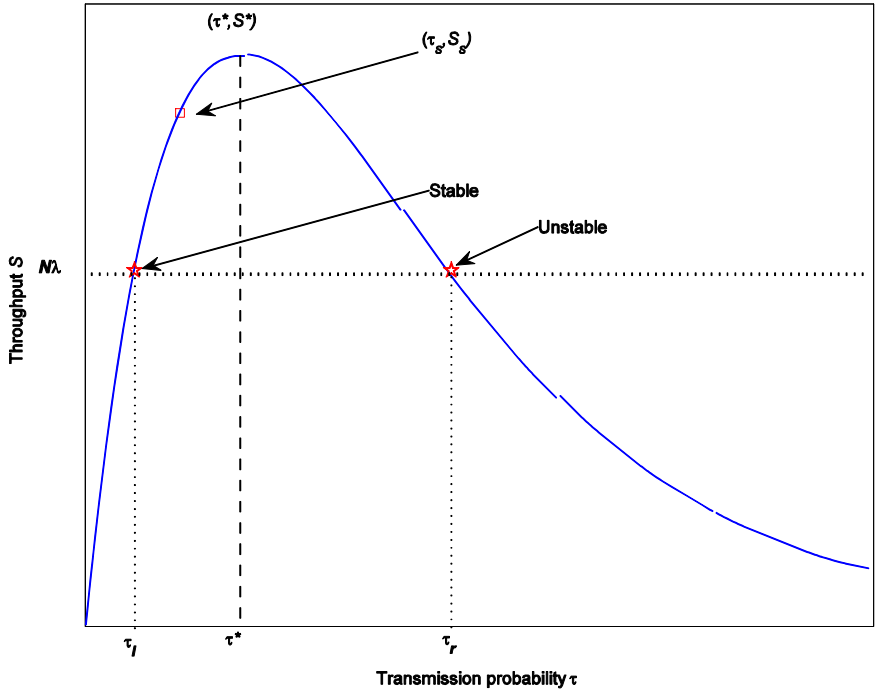

**Fig. 1:** $S-\tau$ curve when $\tau_s < \tau^*$ and $N\lambda < S_s$

By contrast, under non-saturation condition where queues are empty from time to time, throughput is equal to the input traffic rate (also referred to as offered load). That is,

$$N\lambda = \frac{\sum_{k=1}^{M} k \Pr\{X=k\} PL}{P_{idle}^{G} T_{idle} + P_{coll}^{G} T_{coll} + P_{succ}^{G} T_{succ}}. \quad (12)$$

In this case, $\tau$ depends on offered load $N\lambda$ only and is invariant of $r$ and $M$. As shown in Fig. 1, there are two roots to (12), denoted by $\tau_l$ and $\tau_r$, respectively. In particular, there are three cases regarding the relationship among $\tau_l$, $\tau_r$, and $\tau_s$: (i) $\tau_l < \tau_s < \tau_r$ when $N\lambda < S_s$; (ii) $\tau_l < \tau_r < \tau_s$, when $N\lambda \geq S_s$ and $\tau_s > \tau^*$; and (iii) $\tau_s < \tau_l < \tau_r$, when $N\lambda \geq S_s$ and $\tau_s < \tau^*$.

Throughout the analysis in this paper, we assume that the system has reached a steady state. That is, the Markov process that models the queueing behavior of the system is positive recurrent [17]. As indicated by Theorem 1, $\tau_r$ in case (i) and both $\tau_l$ and $\tau_r$ in case (iii) cannot be operating points under steady state.

***Theorem 1:*** Any transmission probability $\tau$ higher than $\tau_s$ cannot be an operating point in WLANs under steady state.

The proof is given in Appendix I. In the proof, we make use of some results that are derived in later sections. Readers are suggested to read the proof later after going through Section III.

Before leaving this section, we would like to emphasize an important underlying assumption in our analysis: a station encounters a constant collision probability $p_c$ when it transmits, regardless of its backoff stage and buffer state. The same assumption has been adopted in most previous papers, e.g., [2], and has been shown to be quite accurate when $N$ is large. We will proceed with this assumption for the time being, and will come back to discuss its implication in some special cases later.

We also note that [8, 12] have adopted the assumption that each station, including the tagged one, transmits at a constant transmission probability $\tau$ in a generic time slot to simplify the analysis. However, such assumption fails to capture the variation in packet delay caused by the time-varying channel-access opportunity due to EB. In this paper, we let the transmission probability of the tagged station vary with its backoff state. By doing so, we are able to characterize the convergence of packet delay and its relationship with the EB scheme deployed.

III. DELAY PERFORMANCE

Packet delay in WLANs is composed of two parts: *waiting time and medium-access delay*. In particular, waiting time denotes the time interval from the arrival of a packet to the instant when the packet becomes a HOL packet in the queue, and medium-access delay denotes the time period from the instant when the packet becomes a HOL packet to the instant at which the packet is successfully transmitted.

There is a strong temptation to model the system by an *M/G/*1 queue with medium-access delay being the service time. Unfortunately, as we will elaborate in subsection III-A and III-B, the distribution of medium-access delay experienced by a packet depends on the buffer state seen by the packet upon its arrival. Consequently, the well-studied *M/G/*1 queueing model, which assumes service time is independent of buffer states, cannot be directly applied. In subsection III-C, we will show that the system can be well represented by an *M/G/*1 queue *with multiple vacations*, referred to as $M/G/1/V_m$. The probability distribution (in terms of transform) of packet delay will be given in subsection III-D, where we also derive close-form expressions for mean packet delay and delay jitter. By the end of this section, we will have conveyed the message that mean packet delay and delay jitter can go to infinity when certain conditions are violated, even if the network is operated far below saturation.

*A. Medium-access delay of packets that arrives at a non-empty queue*

A packet arriving at a non-empty queue becomes a HOL packet immediately after the preceding packet is successfully transmitted (assume the DIFS succeeding the transmission of the previous packet is included in the transmission time). Once it becomes a HOL packet, it starts a backoff process and attempts to access the channel whenever the backoff counter reaches zero. As shown in Fig. 2(a), there are three events that contribute to the medium-access delay: backoff timer countdown, collisions involving the tagged station, and successful transmissions of the tagged station. In particular, the backoff process consists of initial backoff and the backoff periods following unsuccessful transmissions of the tagged station. Given $p_c$ (see (11)), the probability that a packet is successfully transmitted on its $j^{th}$ transmission is given by

$$\Pr\{R = j\} = p_c^{j-1}(1-p_c) \qquad \forall j \geq 1 \qquad (13)$$

where $R$ is the random variable representing the number of transmissions until the successful delivery of the packet. Note that $R$ also denotes the number of backoff periods that contributes to the medium-access delay.

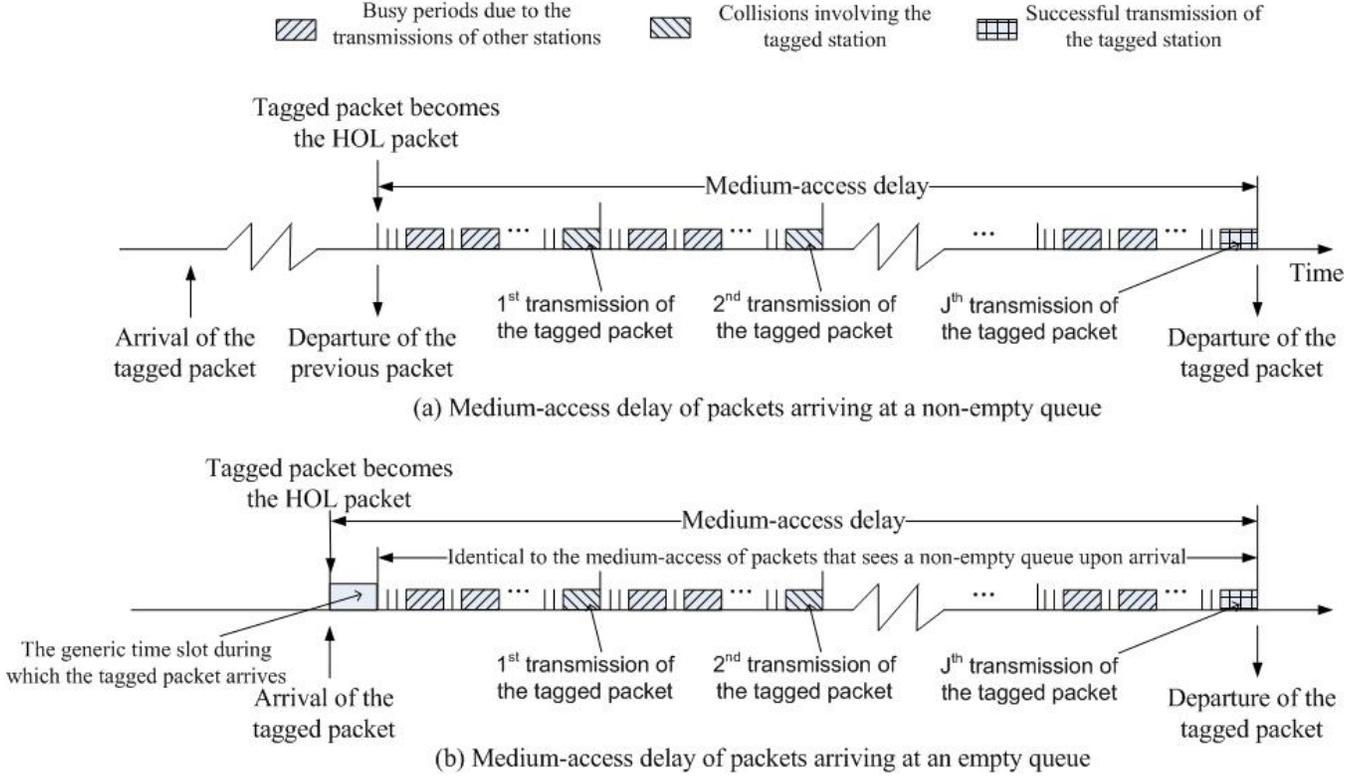

**Fig. 2:** Medium access delay

As mentioned in Section II, the number of countdown slots in the backoff period between the $i-1^{th}$ and the $i^{th}$ transmission, denoted by $B_i$, follows a discrete uniform distribution. That is,

$$\Pr\{B_i = k\} = \begin{cases} \dfrac{1}{r^{i-1}W_0} & k \in [0, r^{i-1}W_0 - 1] \\ 0 & \text{otherwise} \end{cases}, \qquad (14)$$

and the corresponding $Z$ transform is

$$B_i(Z) = \sum_{k=0}^{r^{i-1}W_0 - 1} \Pr\{B_i = k\} z^k = \frac{1 - z^{r^{i-1}W_0}}{r^{i-1}W_0 (1 - z)} \qquad (15)$$

Meanwhile, each countdown slot can be either idle or occupied by collisions and successful transmissions *not* involving the tagged station[1] with the following probabilities: (where superscript $B$ stands for time slots in the "backoff" process of the tagged station)

$$P_{idle}^B = (1 - \tau)^{N-1}, \qquad (16)$$

---

[1] In this paper, we make the same assumption as in [2]. That is, the backoff counter value is reduced by 1 after each idle or busy time slot.

$$P^B_{coll} = 1 - \sum_{k=0}^{M}\binom{N-1}{k}\tau^k(1-\tau)^{N-1-k}, \tag{17}$$

and

$$P^B_{succ} = \sum_{K=1}^{M}\binom{N-1}{k}\tau^k(1-\tau)^{N-1-k}. \tag{18}$$

Let $L$ be a random variable denoting the length of a countdown slot. The Laplace transform of $L$ is given by

$$L^*(s) = \mathrm{E}[e^{-sL}] = e^{-sT_{idle}}P^B_{idle} + e^{-sT_{coll}}P^B_{coll} + e^{-sT_{succ}}P^B_{succ}. \tag{19}$$

The duration of the backoff period between the $i-1^{th}$ and the $i^{th}$ transmission, denoted by $C_i$, can now be calculated as the sum of the lengths of $B_i$ countdown time slots. That is,

$$C_i = \sum_{i=1}^{B_i} L_m, \tag{20}$$

where $L_m$ denotes the $m^{th}$ countdown time slots, which are identically and independently distributed (i.i.d.). The Laplace transform of $C_i$ is calculated as

$$\begin{aligned}C_i^*(s) &= \mathrm{E}[\exp(-sC_i)] \\ &= \sum_{k=0}^{r^{i-1}W_0-1}\mathrm{E}\left[\exp\left(-s\sum_{m=1}^{k}L_m\right)\right]\Pr\{B_i=k\} \\ &= \sum_{k=0}^{r^{i-1}W_0-1}\prod_{m=1}^{k}\mathrm{E}[\exp(-sL_m)]\Pr\{B_i=k\} \\ &= \sum_{k=0}^{r^{i-1}W_0-1}\left(L^*(s)\right)^k \Pr\{B_i=k\} \\ &= B_i(L^*(s)),\end{aligned} \tag{21}$$

from which we derive mean, variance, and the third moment around the mean of $C_i$ as follows.

$$\mathrm{E}[C_i] = -\frac{dC_i^*(s)}{ds}\bigg|_{s=0} = A_1 \frac{r^{i-1}W_0-1}{2}, \tag{22}$$

$$\begin{aligned}\mathrm{VAR}[C_i] &= \frac{d^2C_i^*(s)}{ds^2}\bigg|_{s=0} - \mathrm{E}[C_i]^2 \\ &= A_1^2\frac{(r^{i-1}W_0-1)(r^{i-1}W_0-5)}{12} + A_2\frac{r^{i-1}W_0-1}{2},\end{aligned} \tag{23}$$

and

$$M_3(C_i) = \mathrm{E}\left[(C_i - \mathrm{E}[C_i])^3\right]$$
$$= \frac{A_1^3}{4}\left(-(r^{i-1}W_0)^2 + 4r^{i-1}W_0 - 3\right) + \frac{A_1 A_2}{4}\left((r^{i-1}W_0)^2 - 6r^{i-1}W_0 + 5\right) + A_3 \frac{r^{i-1}W_0 - 1}{2}, \quad (24)$$

where constant $A_1 = T_{succ} P_{succ}^B + T_{coll} P_{coll}^B + T_{idle} P_{idle}^B$, $A_2 = T_{succ}^2 P_{succ}^B + T_{coll}^2 P_{coll}^B + T_{idle}^2 P_{idle}^B$, and $A_3 = T_{succ}^3 P_{succ}^B + T_{coll}^3 P_{coll}^B + T_{idle}^3 P_{idle}^B$.

We are now ready to derive the medium-access delay of packets that arrive at a non-empty queue, denoted by $X_{ne}$, as follows.

$$X_{ne} = \sum_{i=1}^{j} C_i + (j-1)T_{coll} + T_{succ} \quad \text{if } R = j, \quad (25)$$

with Laplace transform being

$$\begin{aligned}
X_{ne}^*(s) &= \sum_{j=1}^{\infty} \mathrm{E}\left[\exp\left(-s\left(\sum_{i=1}^{j} C_i + (j-1)T_{coll} + T_{succ}\right)\right)\right] \Pr\{R = j\} \\
&= \sum_{j=1}^{\infty} \prod_{i=1}^{j} \mathrm{E}[\exp(-sC_i)] \exp\left(-s\left((j-1)T_{coll} + T_{succ}\right)\right) \Pr\{R = j\} \\
&= \sum_{j=1}^{\infty} \prod_{i=1}^{j} C_i^*(s) \exp\left(-s\left((j-1)T_{coll} + T_{succ}\right)\right) \Pr\{R = j\} \\
&= \sum_{j=1}^{\infty} \prod_{i=1}^{j} B_i\left(L^*(s)\right) \exp\left(-s\left((j-1)T_{coll} + T_{succ}\right)\right) \Pr\{R = j\}
\end{aligned} \quad (26)$$

where the second equality is due to the fact that $C_i$'s are independent for different $i$'s.

From (25), the first three moments of $X_{ne}$ can be expressed as functions of $\mathrm{E}[C_i]$, $\mathrm{VAR}[C_i]$, and $M_3(C_i)$ as

$$\begin{aligned}
\mathrm{E}[X_{ne}] &= \sum_{j=1}^{\infty} \mathrm{E}[X_{ne} \mid R = j] \Pr\{R = j\} \\
&= \sum_{j=1}^{\infty} \left(\sum_{i=1}^{j} \mathrm{E}[C_i] + (j-1)T_{coll} + T_{succ}\right) \Pr\{R = j\}
\end{aligned} \quad (27)$$

$$E\left[X_{ne}^2\right] = \sum_{j=1}^{\infty} E\left[X_{ne}^2 \mid R = j\right] \Pr\{R = j\}$$

$$= \sum_{j=1}^{\infty} \text{VAR}[X_{ne} \mid R = j] \Pr\{R = j\} + \sum_{j=1}^{\infty} \left(E[X_{ne} \mid R = j]\right)^2 \Pr\{R = j\} \quad (28)$$

$$= \sum_{j=1}^{\infty} \sum_{i=1}^{j} \text{VAR}[C_i] \Pr\{R = j\} + \sum_{j=1}^{\infty} \left(\sum_{i=1}^{j} E[C_i] + (j-1)T_{coll} + T_{succ}\right)^2 \Pr\{R = j\}$$

$$E\left[X_{ne}^3\right] = \sum_{j=1}^{\infty} E\left[X_{ne}^3 \mid R = j\right] \Pr\{R = j\}$$

$$= \sum_{j=1}^{\infty} M_3(X_{ne} \mid R = j) \Pr\{R = j\} + 3\sum_{j=1}^{\infty} E\left[X_{ne}^2 \mid R = j\right] E\left[X_{ne} \mid R = j\right] \Pr\{R = j\}$$

$$-2\sum_{j=1}^{\infty} \left(E[X_{ne} \mid R = j]\right)^3 \Pr\{R = j\}$$

$$= \sum_{j=1}^{\infty} \sum_{i=1}^{j} M_3(C_i) \Pr\{R = j\} + 3\sum_{j=1}^{\infty} \left(\left(\sum_{i=1}^{j} E[C_i] + (j-1)T_{coll} + T_{succ}\right) \sum_{i=1}^{j} \text{VAR}[C_i]\right) \Pr\{R = j\}$$

$$+ \sum_{j=1}^{\infty} \left(\sum_{i=1}^{j} E[C_i] + (j-1)T_{coll} + T_{succ}\right)^3 \Pr\{R = j\} \quad (29)$$

After some tedious but straightforward derivations, we find that $E\left[X_{ne}^n\right]$ is convergent only when $p_c < 1/r^n$ for $n = 1, 2, 3$, respectively. In particular, when $p_c < 1/r$, the summation in (27) converges to:

$$E[X_{ne}] = A_1 \frac{W_0(1-p_c) - (1-rp_c)}{2(1-p_c)(1-rp_c)} + T_{coll} \frac{p_c}{1-p_c} + T_{succ}. \quad (30)$$

When $p_c < 1/r^2$, the summation in (28) converges to:

$$E\left[X_{ne}^2\right] = A_1^2 \left(\frac{W_0^2}{12(1-r^2 p_c)} - \frac{W_0}{(1-rp_c)} + \frac{5}{12(1-p_c)} + \frac{W_0^2(1+rp_c)}{4(1-rp_c)(1-r^2 p_c)} - \frac{W_0^2(1-rp_c^2)}{2(1-p_c)(1-rp_c)^2} + \frac{1+p_c}{4(1-p_c)^2}\right)$$

$$+ \left(A_2 \frac{W_0(1-p_c)-(1-rp_c)}{2(1-rp_c)(1-p_c)} + T_{coll}^2 \frac{p_c(1+p_c)}{(1-p_c)^2} + T_{succ}^2 + 2A_1 T_{coll} p_c \left(\frac{W_0(1+r-2rp_c)}{2(1-p_c)(1-rp_c)^2} - \frac{1}{(1-p_c)^2}\right)\right)$$

$$+ \left(A_1 T_{succ} \frac{W_0(1-p_c)-(1-rp_c)}{(1-rp_c)(1-p_c)} + 2T_{coll} T_{succ} \frac{p_c}{(1-p_c)}\right). \quad (31)$$

Likewise (29) converges when $p_c < 1/r^3$:

$$E\left[X_{ne}^3\right] = \theta_1 + 3\theta_2 + \theta_3, \tag{32}$$

where $\theta_1$, $\theta_2$, and $\theta_3$ are given in Appendix II.

**Remark 1:** In EB schemes where $r > 1$, $p_c < 1/r^{n_1}$ is a tighter condition than $p_c < 1/r^{n_2}$ if $n_1 > n_2$. In other words, the convergence of $E[X_{ne}^{n_1}]$ implies the convergence of $E[X_{ne}^{n_2}]$ for $n_1 > n_2$, but not the reverse.

*B. Medium-access delay of packets that arrive at an empty queue*

Packets that arrive at empty queues undergo a different medium access delay than those derived in last subsection. As shown in Fig. 2(b), a packet that arrives at an empty queue becomes a HOL packet immediately after its arrival. The arrival may occur in the middle of an idle time slot or a time slot that is occupied by collisions or successful transmissions of other stations. The probability of the slot being idle or occupied by collisions and success transmissions is the same as that given in (16)-(18), respectively. According to the protocol, the station cannot access the channel until the end of the time slot during which the packet arrives. Once the time slot ends, the backoff process starts and the packet will be transmitted once the backoff timer counts down to zero[2]. When $N$ is relatively large, the channel states of adjacent time slots are effectively independent of each other. As a result, the backoff process, once started, is stochastically identical to the one described in Section IIIA. That is, the time period between the instant when the backoff process starts and the instant when the packet is successfully transmitted has the same distribution as $X_{ne}$. In other words, the medium-access delay of packets arriving at an empty queue, denoted by $X_e$, consists of two parts: a time period that is statistically identical to $X_{ne}$ and an additional waiting time before the backoff process starts. Therefore,

$$X_e \geq X_{ne}. \tag{33}$$

So far, we have shown that the distribution of medium-access delay depends on the buffer state upon the arrival of a packet. Hence, the well-studied $M/G/1$ queueing model, which assumes service time is independent of buffer states, cannot be directly applied. In the next subsection, we model the system with an $M/G/1/V_m$ queue model. For this model, it is sufficient to know $X_{ne}$ to characterize the queueing performance of the system. We therefore do not derive explicit expressions for $X_e$ here to conserve space.

---

[2]In this paper, we assume a variant of 802.11 DCF, i.e., each packet goes through a backoff process regardless of the system state it sees upon arrival. In some practical systems, a packet is transmitted immediately after the channel is idle for a DIFS time, if the packet arrives at an empty queue and finds the channel is idle. Compared with our system, the mean delay at low traffic load is offset by roughly $(W_0 - 1)/2$, while the delay at high traffic load is about the same. In this paper, we focus on the capacity limit of the system when it is heavily loaded, which is of more theoretical interest. Our assumption captures the fundamental characteristics of the system, while making the analysis much more tractable.

## C. $M/G/1/V_m$ queueing model

As discussed in the above two subsections, all packets of the tagged station experience a service time of $X_{ne}$, when the tagged station is continuously backlogged. Once the station becomes idle, the channel will be occupied by an idle time slot or a busy time slot due to the transmissions of *other* stations. If the tagged station is still empty at the end of the time slot, the channel will be occupied by another time slot that is i.i.d. to the previous one. Otherwise, the tagged station will start a backoff process and its HOL packet will experience a service time of $X_{ne}$.

The queueing behavior described above is well modeled by an $M/G/1/V_m$ queue [16] where the server takes a vacation every time the system becomes empty. In our case, the vacation period has the same distribution as in (19). Hence, the forward recurrence time of vacation, denoted by $Y$, has a Laplace transform of

$$Y^*(s) = \frac{1-L^*(s)}{s\mathrm{E}[L]} = \frac{1-\left(e^{-sT_{idle}}P^B_{idle} + e^{-sT_{coll}}P^B_{coll} + e^{-sT_{succ}}P^B_{succ}\right)}{S\left(T_{idle}P^B_{idle} + T_{coll}P^B_{coll} + T_{succ}P^B_{succ}\right)}. \tag{34}$$

It can be easily shown that $\mathrm{E}[Y] = \frac{A_2}{2A_1}$ and $\mathrm{E}[Y^2] = \frac{A_3}{3A_1}$.

## D. Packet delay distribution, mean packet delay, and delay jitter

We are now ready to calculate packet delay of WLAN, denoted by $D$, as the system time of an $M/G/1/V_m$ queue with arrival rate $\lambda$, service time $X_{ne}$, vacation time $L$ with forward recurrence time $Y$.

To analyze $M/G/1/V_m$, we first study the performance of the corresponding $M/G/1$ queue without vacation but with the same arrival rate and service time. Let $\tilde{\rho} = \lambda\mathrm{E}[X_{ne}]$ be the utilization of the server (also referred to as HOL occupancy) in the $M/G/1$ queue, and denote $\tilde{D}$ the corresponding system time. According to Pollaczek-Khinchin (P-K) transform equation of $M/G/1$ queue,

$$\tilde{D}^*(s) = X_{ne}^*(s)\frac{s(1-\tilde{\rho})}{\lambda X_{ne}^*(s) - \lambda + s}, \qquad (35)$$

while the number of packets left behind by a departure of a packet has a $Z$ transform of $\tilde{D}^*(\lambda - \lambda z)$ [17].

In the $M/G/1/V_m$ queue, packets are expected to experience a longer delay due to the additional vacation time. In particular, the number of packets that arrive during the forward recurrence time of a vacation is a random variable with $Z$ transform $Y^*(\lambda - \lambda z)$. In our system, the vacation sequence is stationary and the length of a vacation does not depend on the inter-arrival and service time after the end of this vacation. Consequently, the decomposition property as follows applies --- the number of packets left behind by a departure of a packet, denoted by $Q$, is distributed as the sum of two independent random variables: the number of packets at a service completion in the $M/G/1$ queue without vacation and the number of packets that arrive during the forward recurrence time of a vacation. As a result,

$$\hat{Q}(z) = \tilde{D}^*(\lambda - \lambda z)Y^*(\lambda - \lambda z) = (1-\tilde{\rho})X_{ne}^*(\lambda - \lambda z)\frac{(1-z)Y^*(\lambda - \lambda z)}{X_{ne}^*(\lambda - \lambda z) - z}, \qquad (36)$$

It is then straightforward that the packet delay follows a distribution of [16]

$$D^*(s) = \hat{Q}(1-s/\lambda) = (1-\tilde{\rho})X_{ne}^*(s)\frac{sY^*(s)}{\lambda X_{ne}^*(s) - \lambda + s}. \qquad (37)$$

The mean packet delay $E[D]$ and delay jitter $VAR[D]$ can be calculated as

$$E[D] = -\frac{dD^*(s)}{ds}\bigg|_{s=0} = E[X_{ne}] + E[Y] + \frac{\lambda E[X_{ne}^2]}{2(1-\tilde{\rho})}, \qquad (38)$$

$$\begin{aligned}VAR[D] &= E[D^2] - (E[D])^2 = \frac{d^2 D^*(s)}{ds^2}\bigg|_{s=0} - (E[D])^2 \\ &= VAR[X_{ne}] + VAR[Y] + \frac{\lambda^2 \left(E[X_{ne}^2]\right)^2}{4(1-\tilde{\rho})^2} + \frac{\lambda E[X_{ne}^3]}{3(1-\tilde{\rho})}.\end{aligned} \qquad (39)$$

The utilization of the server in the $M/G/1/V_m$ system, denoted by $\rho$, is given by

$$\begin{aligned}\rho &= 1 - \Pr\{Q=0\} = 1 - \hat{Q}(0) \\ &= 1 - (1-\tilde{\rho})Y^*(\lambda) \\ &= 1 - (1-\tilde{\rho})\frac{1 - \left(e^{-\lambda T_{idle}}P_{idle}^B + e^{-\lambda T_{coll}}P_{coll}^B + e^{-\lambda T_{succ}}P_{succ}^B\right)}{\lambda\left(T_{idle}P_{idle}^B + T_{coll}P_{coll}^B + T_{succ}P_{succ}^B\right)}.\end{aligned} \qquad (40)$$

According to queueing theory, a Markov chain associated with a queue can reach a steady state if and only if $\rho < 1$ [17]. It is not difficult to see from (40) that $\rho \geq \tilde{\rho}$ in our system, while $\rho < 1$ if and only if $\tilde{\rho} < 1$. When the system is saturated, $\rho = \tilde{\rho} = 1$. Considering (27), $\tilde{\rho} = \lambda E[X_{ne}] < 1$ can be equivalently written as

$$p_c r + \frac{\lambda A_1 W_0 (1-p_c)}{2(1-p_c) - 2\lambda T_{succ}(1-p_c) - 2\lambda T_{coll} p_c + \lambda A_1} < 1. \tag{41}$$

***Theorem 2:*** Mean packet delay $E[D]$ is finite if and only if $\tilde{\rho} < 1$ and $p_c < 1/r^2$. Likewise, delay jitter $VAR[D]$ is finite if and only if $\tilde{\rho} < 1$ and $p_c < 1/r^3$.

Proof: The proof is straightforward from (38), (39) and the conditions for convergence of the first three moments of $X_{ne}$. Note that $\tilde{\rho} = \lambda E[X_{ne}] < 1$ implies $E[X_{ne}]$ is finite. In addition, $E[X_{ne}^2]$ and $E[X_{ne}^3]$ converge if and only if $p_c < 1/r^2$ and $p_c < 1/r^3$ respectively. When $N$ is very large, $\lambda \to 0$ because system throughput $N\lambda$ is finite. In this case, the second term in the left hand side of (41) converges to 0 and inequality (41) becomes $p_c < 1/r$, which is automatically satisfied when $p_c < 1/r^2$ (or $p_c < 1/r^3$). Therefore, for large $N$, $p_c < 1/r^2$ is the sufficient and necessary condition for finite mean delay $E[D]$, while $p_c < 1/r^3$ is the sufficient and necessary condition for finite delay jitter $VAR[D]$. ∎

***Remark 2:*** It is obvious from Remark 1 and Theorem 2 that mean delay $E[D]$ can be infinite even if the system is in a steady state. Likewise, finite mean delay does not imply finite delay jitter.

Readers are now ready to read the proof of Theorem 1, which is given in Appendix I.

***Remark 3:*** *The proof of Theorem 1 suggests that $\tilde{\rho} < 1$ when $\tau < \tau_s$, since $\tilde{\rho}$ is an increasing function of $\tau$. That is, the system is operated under non-saturation condition whenever the average transmission rate $\tau$ is smaller than $\tau_s$. An interesting scenario arises when $\tau_s > \tau^*$ and $N\lambda > S_s$. In this scenario, both $\tau_l$ and $\tau_r$ are smaller than $\tau_s$. This implies that the system can, in theory, achieve throughput higher than saturation throughput $S_s$, while observing empty queues from time to time. However, as we will show in the next section, it is not safe to load the system with an offered load higher than $S_s$ in practice, because bounded packet delay cannot be sustained for a long time.*

IV. SBMD AND SBDJ THROUGHPUTS

It is generally accepted that a traffic load is sustainable as long as it is lower than saturation throughput. However, Remark 2 reveals the fact that packets may suffer from very large mean delay or delay jitter even when non-zero throughput can be sustained. In many applications, it is crucial to guarantee bounded mean delay or delay jitter. To bridge the gap, we will define in this section SBMD and SBDJ throughputs, which are the highest throughputs that can be sustained with bounded mean delay and delay jitter, respectively.

*A. Boundary-bounded-mean-delay (BBMD) and boundary-bounded-delay-jitter (BBDJ) throughput*

For large $N$, $p_c < 1/r^2$ and $p_c < 1/r^3$ are sufficient and necessary conditions for bounded mean delay and bounded delay jitter respectively. By observing the boundary cases where $p_c = 1/r^2$ and $p_c = 1/r^3$, we can get the *highest* possible transmission probabilities that do not cause unbounded mean delay and bounded delay jitter, respectively. Denote such transmission probabilities by $\tau_{BBMD}$ and $\tau_{BBDJ}$, respectively, and the corresponding throughput by $S_{BBMD}$ and $S_{BBDJ}$. It is obvious from (11) that $p_c$ is a increasing function of $\tau$. Hence,

$$\tau_{BBDJ} < \tau_{BBMD} < \tau_s \tag{42}$$

As discussed in Section II, depending on $r$, $\tau_s$ can be smaller than, equal to, or larger than $\tau^*$. The relationship between $S_s$, $S_{BBMD}$, and $S_{BBDJ}$ highly depends on the position of $\tau_s$. To show this, we illustrate four different scenarios in Fig. 3: $\tau_s \leq \tau^*$ in scenario 1 and $\tau_s > \tau^*$ in the other 3 scenarios. For simple illustration, we focus on $S_{BBMD}$ only. However, the following conclusions can be easily extended to $S_{BBDJ}$ by replacing the inequality $p_c < 1/r^2$ with $p_c < 1/r^3$.

In Fig. 3, the thickened parts of the curves denote the region in which $p_c < 1/r^2$. Operating regions beyond the thickened part in each of the scenarios is not viable if bounded mean delay is to be achieved. In scenarios 1, 2, and 3, mean packet delay becomes unbounded when the input traffic load $N\lambda$ is higher than $S_{BBMD}$. In particular in scenarios 1 and 2 where $S_{BBMD} < S_s$, it is necessary to load the system below the saturation point to avoid excessively long packet delay.

In scenarios 3 and 4, $S_{BBMD} > S_s$. In these cases, it is *theoretically* possible to operate the system at a higher throughput than the saturation throughput $S_s$ while achieving a bounded mean delay. More interestingly, in scenario 4, it is even possible to load the system at the maximum throughput $S^*$ while having a finite mean delay *in theory*, as long as the system is operated within the thickened region of the curve. However, as we will show in the next subsection, it is not safe to load the system with an offered load higher than $S_s$.

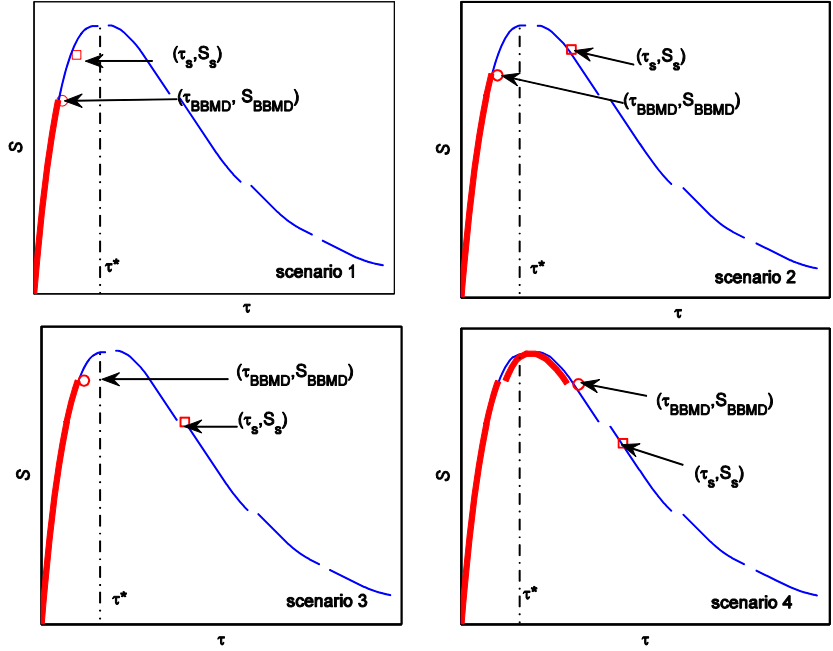

**Fig. 3:** Scenario 1: $\tau_{BBMD} < \tau_s < \tau^*$, $S_{BBMD} < S_s$; Scenario 2: $\tau_{BBMD} < \tau^* < \tau_s$, $S_{BBMD} < S_s$; Scenario 3: $\tau_{BBMD} < \tau^* < \tau_s$, $S_{BBMD} > S_s$; Scenario 4: $\tau^* < \tau_{BBMD} < \tau_s$, $S_{BBMD} > S_s$.

So far, we have discussed the large $N$ case. When $N$ is small to the extent that inequality $p_c < 1/r^2$ (resp. $p_c < 1/r^3$) can always be satisfied as long as the system is not saturated, $\tilde{\rho} < 1$ becomes a stricter condition that $p_c < 1/r^2$ (resp. $p_c < 1/r^3$). In this case, $\tau_{BBMD}$ (resp. $\tau_{BBDJ}$) $= \tau_s$ and $S_{BBMD}$ (resp. $S_{BBDJ}$) $= S_s$. Since the behavior of MPR WLANs at $(\tau_s, S_s)$ has been extensively studied in another paper of ours [14], we focus our interest in the large $N$ case in this paper.

*B. SBMD and SBDJ throughputs*

Scenarios 3 and 4 in Fig. 3 imply that it is theoretically possible to operate the system at a throughput higher than $S_s$ while maintaining $p_c < 1/r^2$ (or $p_c < 1/r^3$ for BDJ). This is the case only if the long-term average output rate can be maintained at the higher throughput. In practice, however, it is not safe to load the system with an input rate higher than $S_s$. To see this, we note that because of the random nature of the system, the ``instantaneous'' input rate and output rate of the whole system consisting of all queues may vary over time. It is possible for the system to evolve to a state where the instantaneous input rate is larger than the instantaneous output rate, and therefore for the backlog at the queues to build up. If this persists for a while, the system (all queues) may become saturated and the "long-term" output rate will then degenerate to the saturation throughput, which is lower than the original higher long-term output rate. If the "long-term" average input rate, the offered load $N\lambda$, is set at above the saturation throughput, the backlog will continue to build up and the system will not get out of saturation. The delay will then go to infinity. The intricacy lies in the fact that we have a system in which the service rate can be degraded ($p_c$ increases) once saturation sets it, which is unlike an "ordinary" queuing system in which the service rate is independent of the system state. This particular aspect of $p_c$ transiting to a higher value under saturation is not captured in our $M/G/1/V_m$ queue model, which looks at a tagged queue and assumes a constant $p_c$. This phenomenon will be illustrated by numerical simulations in the Section V.

Define safe BMD throughput $S_{SBMD}$ and safe BDJ throughput $S_{SBDJ}$ to be the highest throughput that can be *safely* sustained with bounded mean delay and delay jitter, respectively. Based on the above articulation,

$$S_{SBMD} = \min(S_{SBMD}, S_s) \tag{43}$$

and
$$S_{SBDJ} = \min(S_{SBDJ}, S_s) \tag{44}$$

*C. Super-linear scaling of maximum SBMD and SBDJ throughput*

In our earlier work [10, 11, 14], we have proved that the maximum saturation throughput of MPR WLANs increases *super-linearly* with MPR capability $M$. In this subsection, we will show that super-linear scaling also holds for the maximum SBMD and SBDJ throughput. That is $\frac{S^*_{SBMD}}{M}$ and $\frac{S^*_{SBDJ}}{M}$ increases with $M$.

Given $M$, $S_{SBMD}$ and $S_{SBDJ}$ can be maximized by deploying an optimal $r$ in the EB process. Denote the maximum $S_{SBMD}$ and $S_{SBDJ}$ by $S^*_{SBMD}(M)$ and $S^*_{SBDJ}(M)$, respectively. More precisely,

$$S^*_{SBMD}(M) = S_{SBMD}(M, r^*_{SBMD}(M)) \tag{45}$$

and
$$S^*_{SBDJ}(M) = S_{SBDJ}(M, r^*_{SBDJ}(M)) \tag{46}$$

where $r^*_{SBMD}(M)$ and $r^*_{SBDJ}(M)$ denote the optimal backoff factors that maximize $S^*_{SBMD}(M)$ and $S^*_{SBDJ}(M)$.

*The large N case*

From (43), the optimal $r$ that maximizes $S_{SBMD}$ is such that

$$S_{BBMD}(M, r^*_{SBMD}(M)) = S_s(M, r^*_{SBMD}(M)) \tag{47}$$

As $\tau_{BBMD} < \tau_s$ in the large $N$ case, $r^*_{SBMD}(M)$ that satisfies (47) yields $\tau_{BBMD} < \tau^*$ and $\tau_s > \tau^*$ (i.e., somewhere between scenario 2 and scenario 3 defined in Fig. 3).

To solve (47), we note that the number of attempts in a backoff slot can be approximated by a Poisson process with an average attempt rate $\eta = N\tau$ when $N$ is relatively large. That is,

$$\Pr\{X = k\} = \frac{\eta^k e^{-\eta}}{k!} \tag{48}$$

Substituting (48) into (47), we have

$$\frac{PL\eta_{BBMD} \sum_{k=1}^{M-1} \frac{\eta_{BBMD}^k e^{-\eta_{BBMD}}}{k!}}{P^G_{idle}(\eta_{BBMD})T_{idle} + P^G_{coll}(\eta_{BBMD})T_{coll} + P^G_{succ}(\eta_{BBMD})T_{succ}} = \frac{PL\eta_s \sum_{k=1}^{M-1} \frac{\eta_s^k e^{-\eta_s}}{k!}}{P^G_{idle}(\eta_s)T_{idle} + P^G_{coll}(\eta_s)T_{coll} + P^G_{succ}(\eta_s)T_{succ}} \tag{49}$$

For large $N$, $p_c \to 1/r$ when $\eta \to \eta_s$ and $p_c \to 1/r^2$ when $\eta \to \eta_{BBMD}$. That is,

$$\sum_{k=1}^{M-1} \frac{\eta_s^k e^{-\eta_s}}{k!} = 1 - p_c = 1 - 1/r, \tag{50}$$

and

$$\sum_{k=1}^{M-1} \frac{\eta_{BBMD}^k e^{-\eta_{BBMD}}}{k!} = 1 - p_c = 1 - 1/r^2, \tag{51}$$

From (49)--(51), $r^*_{SBMD}(M)$ and $S^*_{SBMD}(M)$ can be solved numerically. Likewise, we can solve for $r^*_{SBDJ}$ and $S^*_{SBDJ}$ by replacing $1/r^2$ with $1/r^3$ in the above equations.

In Fig. 4, $\frac{S^*_{SBMD}}{M}$ and $\frac{S^*_{SBDJ}}{M}$ are plotted against $M$ for slotted ALOHA, with the corresponding $r^*_{SBMD}(M)$ and $r^*_{SBDJ}(M)$ plotted in Fig. 5. Detailed parameters and values are listed in Table I. The figure shows that SBMD and SBDJ throughput scale super-linearly with $M$: normalized throughput $\frac{S^*_{SBMD}}{M}$ and $\frac{S^*_{SBDJ}}{M}$ increase with $M$. This result, together with our earlier work [14], provides a strong incentive to deploy MPR in future WLANs, no matter whether the underlying application is delay sensitive or not.

**Table I:** System Parameters

| Parameter | Value |
|---|---|
| PHY Header | 20 $\mu s$ |
| MAC Header | 244 bits transmitted at 6 Mbps |
| Data Transmission Rate | 6 Mbps |
| CWmin | 16 |
| CWmax | Inf |
| Retry Limit | Inf |
| DIFS | 34 $\mu s$ |
| SIFS | 16 $\mu s$ |

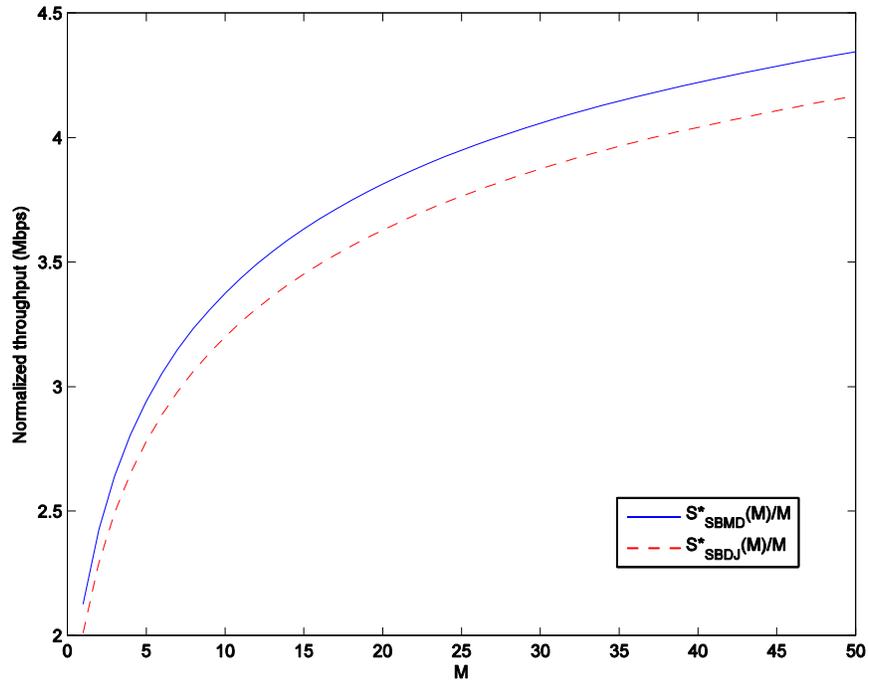

**Fig. 4:** Super-linear throughput scaling in ALOHA network.

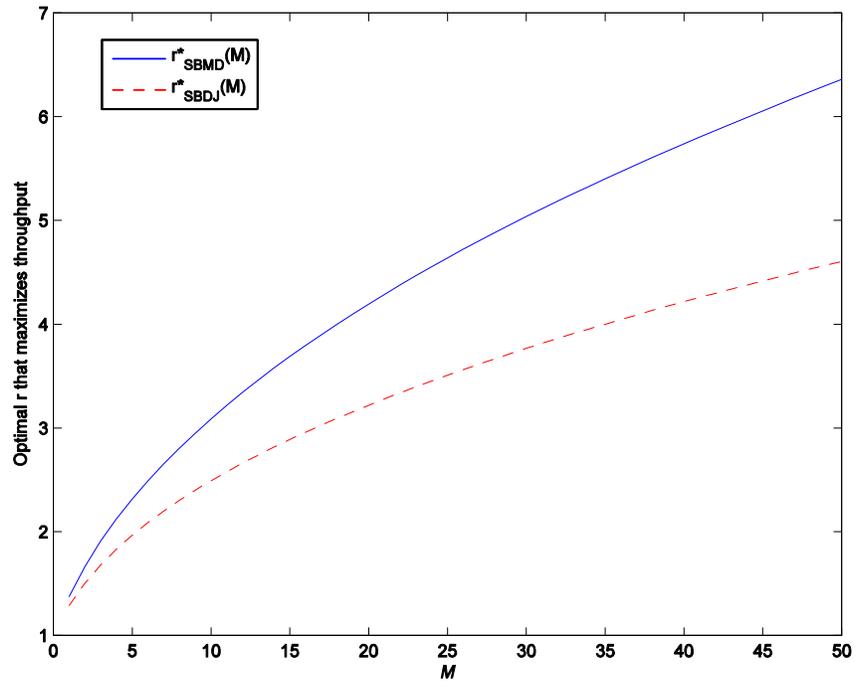

**Fig. 5:** Optimal *r* that maximizes SBMD and SBDJ throughputs in ALOHA network.

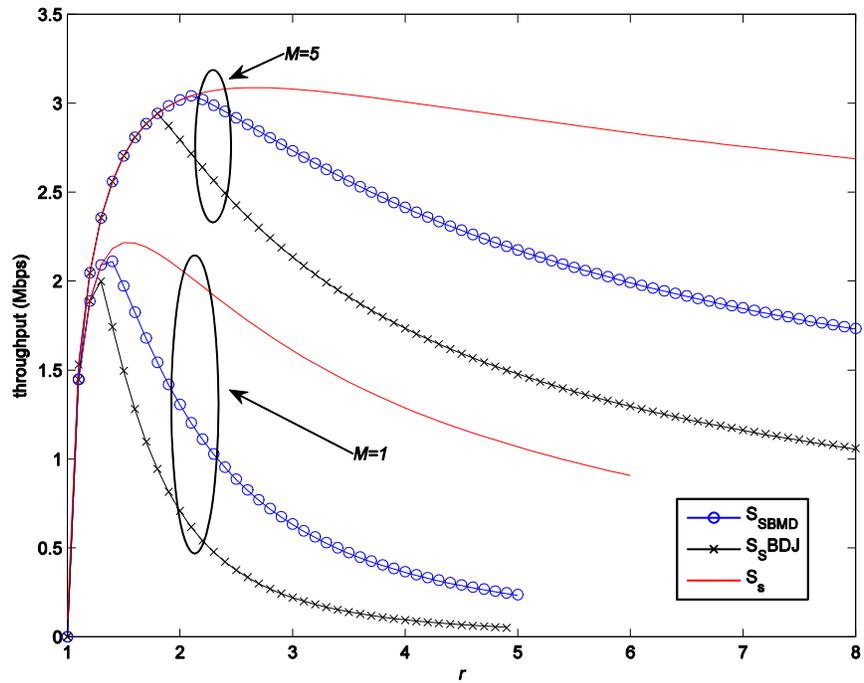

**Fig. 6:** Throughput vs. *r* in ALOHA network.

It is worth noting that SBMD and SBDJ throughputs are more sensitive to $r$ than saturation throughput, as shown in Fig. 6. Therefore, one should be more careful in choosing the right $r$ to avoid severe degradation in sustainable throughput when delay is a concern. In addition, depending on $M$, the commonly adopted binary EB, where $r=2$, can be far from optimum. As Fig. 5 indicates, the optimal $r$'s that achieve $S^*_{SBMD}(M)$ and $S^*_{SBDJ}(M)$ increase with $M$.

A close observation of Fig. 6 shows that large MPR capability $M$ decreases the sensitive of throughput to $r$. This provides another incentive to deploy MPR in WLANs, because the system is now more robust against mis-selection of $r$.

Super-linear throughput scaling is also observed in carrier-sensing networks. In Fig. 7 and Fig. 8, throughput and normalized throughput are plotted for DCF basic-access and RTS/CTS access modes, respectively. The figures show that SBMD and SBDJ throughputs are greatly improved due to the MPR enhancement in the PHY layer. Moreover, $\frac{S^*_{SBMD}}{M}$ and $\frac{S^*_{SBDJ}}{M}$ increase with $M$ for both access modes when $M$ is relatively large.

*The small $N$ case*

When $N$ is small to the extent that $(\tau_{BBMD}, S_{BBMD})$ and/or $(\tau_{BBDJ}, S_{BBDJ})$ overlap with $(\tau_s, S_s)$, $S^*_{SBMD}(M)$ or $S^*_{SBDJ}(M)$ are equal to $S^*(M)$, the maximal saturation throughput. In our earlier work in [14] we have proved that $S^*(M)$ scales super-linearly with $M$. Hence, super-linear scaling of $S^*_{SBMD}(M)$ or $S^*_{SBDJ}(M)$ is straightforward in this case.

## V. NUMERICAL RESULTS

In this section, we further illustrate the results in Section III and Section IV through two examples: slotted ALOHA and DCF basic access systems with binary EB and $M=1$ and $N=50$. Other system parameters are listed in Table I. A simple calculation shows that $\tau_{BBDJ} < \tau_{BBMD} < \tau_s \leq \tau^*$ and $S_{BBDJ} < S_{BBMD} \leq S_s$ in the slotted ALOHA system, while $\tau^* < \tau_{BBDJ} < \tau_{BBMD} < \tau_s$ and $S_s < S_{BBDJ} < S_{BBMD}$ in the DCF system with basic access mode. That is, the two example systems fall in scenario 1 and 4 as defined in Fig. 3, respectively.

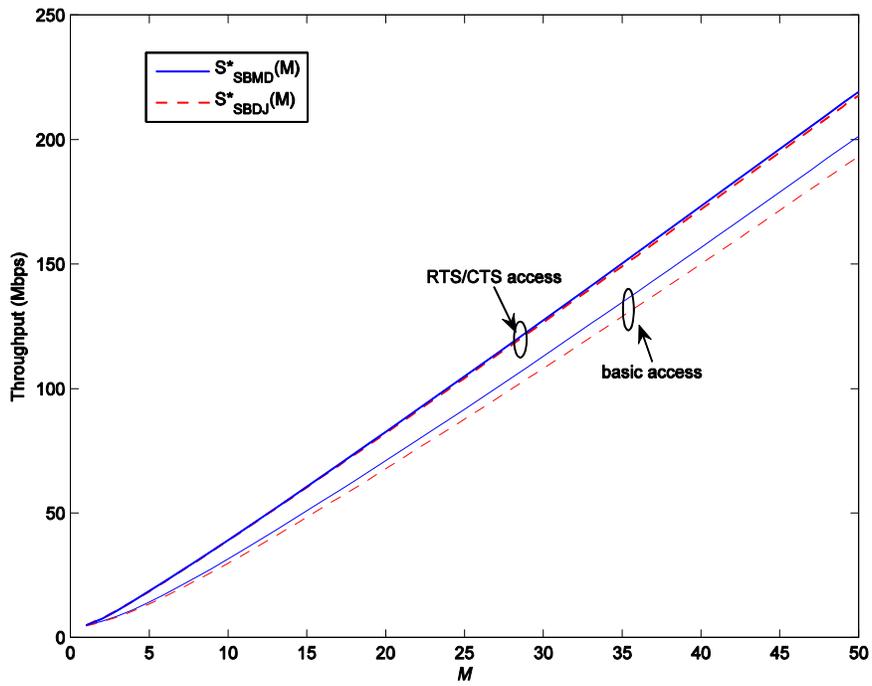

**Fig. 7:** Throughput vs. $M$ in carrier-sensing networks.

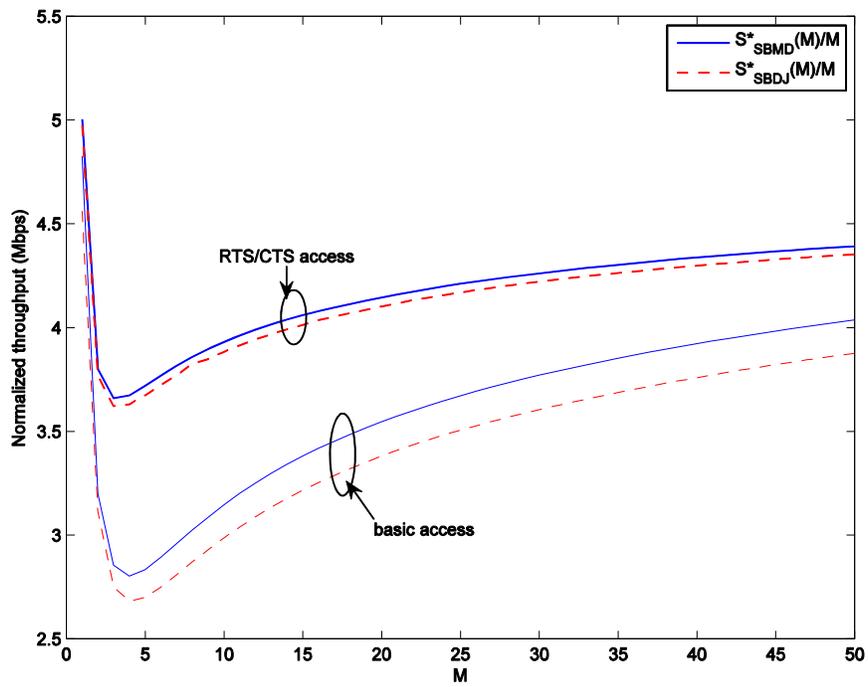

**Fig. 8:** Super-linear throughput scaling in carrier-sensing networks

*Slotted ALOHA System: Scenario 1*

In this scenario, it is only possible to operate the system at an offered load lower than $S_{BBMD}$, and there is only one possible operating point, $\tau_l$, for each offered load $N\lambda$.

In Fig. 9, the utilization factor $\rho$ is plotted against offer load $N\lambda$. Unlike traditional queueing systems where $\rho$ increases at the same rate as offered load, the figure shows that $\rho$ increases much faster than $N\lambda$ in WLANs, especially when $N\lambda$ is large. This is due to the fact that as $\lambda$ increases, not only does the input rate of the tagged queue increases, the mean service time $E[X_{ne}]$ also increases due to heavier contention among nodes. Beyond certain point, the system approaches saturation (i.e., $\rho=1$) very rapidly.

In Fig. 10, we plot $E[X_{ne}]$, $E[D]$, and $\sigma[D]=\sqrt{VAR[D]}$ against offered load $N\lambda$. The solid lines represent the results obtained from analysis. The markers correspond to simulation results. It is not surprising that when the offered load reaches the saturation throughput (which is also the point at which $\rho$ goes to 1 in Fig. 9), $E[X_{ne}]$ quickly converges to a constant equal to the reciprocal of the saturation throughput of one user. As predicted by the analysis, mean packet delay $E[D]$ becomes infinite earlier than $E[X_{ne}]$ because $S_{BBMD} < S_s$. Likewise, the offered load that can be sustained with finite delay jitter is even lower: $\sigma[D]$ approaches infinity earlier than $E[D]$. In this scenario, it is necessary to load the system far below the saturation throughput to guarantee finite delay and delay jitter.

In this figure, we have conducted several independent simulation experiments to measure packet delay. One interesting observation is that different simulation experiments do not yield the same results when offered load is relatively high, even if we run each experiment for a long time (at the order of hours) This is, however, not surprising. When offered load is higher than $S_{BBMD}$, $E[X_{ne}^2]$ is infinite, and so is $VAR[X_{ne}]$. Hence, sample mean of $X_{ne}$ obtained from numerical simulation will not converge to the true mean $E[X_{ne}]$ no matter how much data are collected. Likewise, when $\sigma[D]$ is infinite, the simulation results for mean packet delay do not converge. Interested readers are referred to [18], where we discuss this issue in more depth.

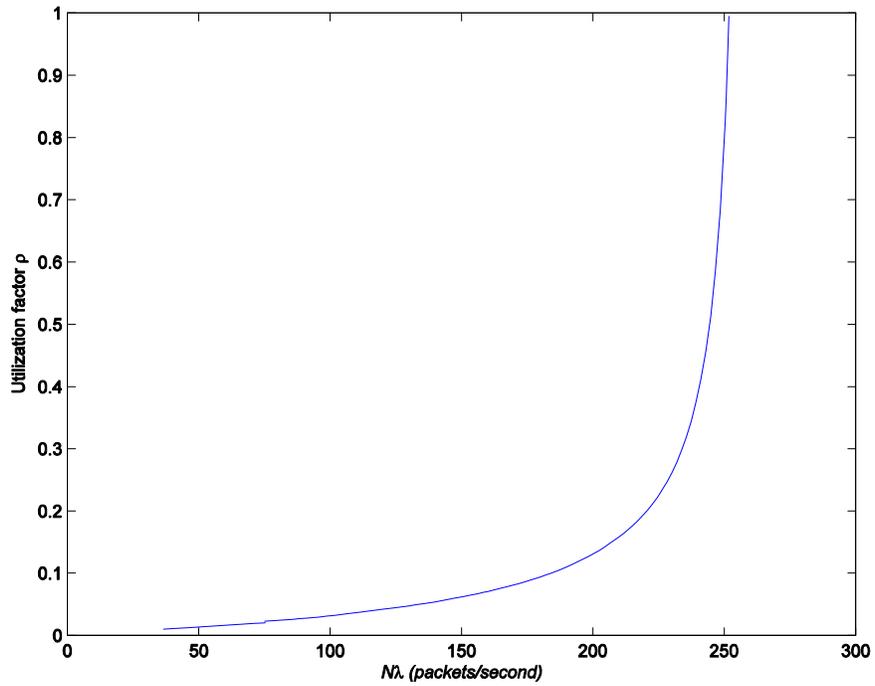

**Fig. 9:** Utilization factor vs. offered load in ALOHA networks

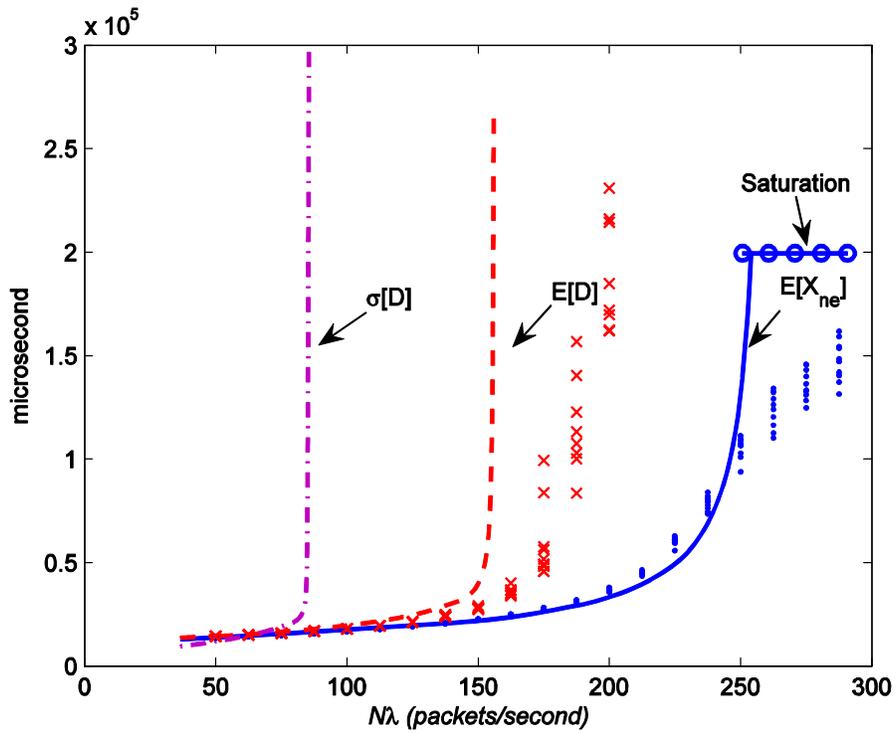

**Fig. 10:** Delay vs. offered in ALOHA networks

*DCF System with Basic Access Mode: Scenario 4*

Similar to scenario 1, $\tau_l$ is the only operating point in scenario 4 when $N\lambda < S_s$. When $N\lambda > S_s$, however, both $\tau_l$ and $\tau_r$ are smaller than $\tau_s$. In other words, an offered load $S_s < N\lambda < S^*$ can result in two attempt rates under non-saturation condition. $\tau_l$ corresponds to a lower contention level, while $\tau_r$ leads to a higher contention level. This is illustrated in Fig. 11, where $\rho$ is plotted against $N\lambda$. It can be seen that when $N\lambda$ is larger than $S_s$, which is around 486.5 packets per second, there are two $\rho$'s corresponding one $N\lambda$. The smaller $\rho$ results from $\tau_l$ and the larger one results from $\tau_r$. If the system operates at $\tau_r$, it reaches saturation when $N\lambda$ approaches $S_s$.

In Fig. 12, $E[X_{ne}]$, $E[D]$, and $\sigma[D] = \sqrt{VAR[D]}$ are plotted against $N\lambda$. The curves without marks represent the results obtained from analysis. Similar to the arguments in Fig. 11, there are two $E[X_{ne}]$'s corresponding to one $N\lambda$ when $N\lambda > S_s$. When saturated, $E[X_{ne}]$ is equal to the reciprocal of the saturation throughput of one station. Likewise, ``kinks'' in $E[D]$ and $\sigma[D]$ are also observed when $N\lambda > S_{BBMD}$ and $N\lambda > S_{BBDJ}$, respectively. (Note that the ``kink'' in $\sigma[D]$ is not obvious because $S_{BBDJ}$ is close to $S^*$ in this example).

As discussed in Section IV.B, it is not safe to load the system with an offered load higher than $S_s$ in practice. Otherwise, system throughput will eventually collapse to $S_s$ and packet delay will go to infinity. To see this, we over-plot the simulation results in Fig. 12. As expected, we are unable observe a throughput higher than $S_s$ in the simulations. When offered load $N\lambda$ approaches $S_s$, the mean service time quickly converges to the reciprocal of saturation throughput, implying that the system is already saturated. In the meantime, packet delay becomes unbounded as well.

Note that unlike in the ALOHA case, numerical results from different simulation runs converge in the basic-access mode. This is because for the region of bounded mean delay, the variance of delay does not go to infinity.

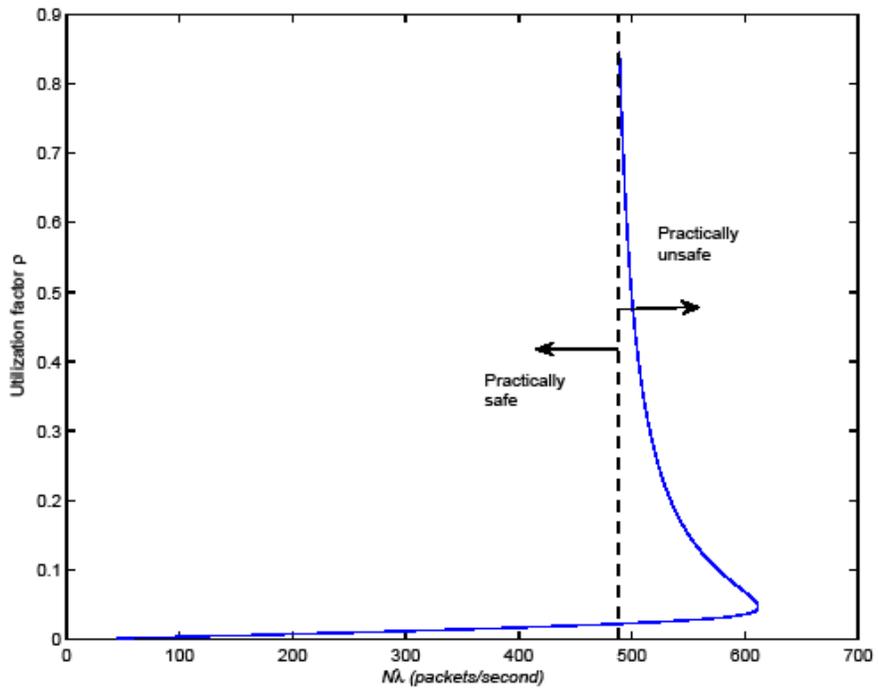

**Fig. 11:** Utilization factor vs. offered load in WLANs with basic-access mode

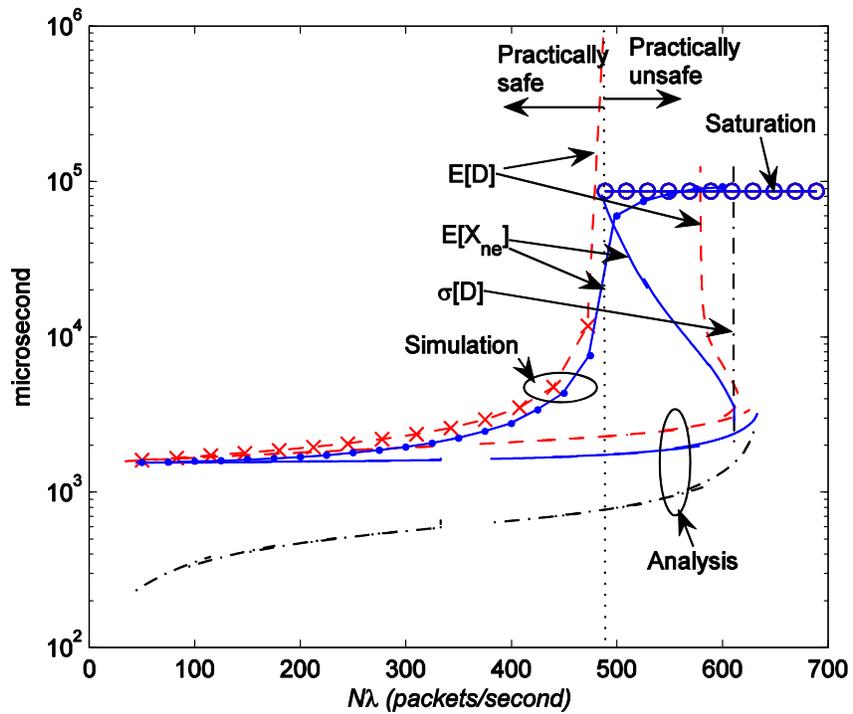

**Fig. 12:** Delay vs. offered load in WLANs with basic-access networks

## VI. CONCLUSIONS AND DISCUSSIONS

In this paper, we have investigated the delay performance of EB-based WLANs with MPR capability under non-saturation condition. Using an $M/G/1/V_m$ queueing model, we have derived an explicit expression for the distribution (in transform) of packet delay. The analysis establishes sufficient and necessary conditions for mean delay and delay jitter to be bounded: $\tilde{\rho} < 1$ and $p_c < 1/r^2$ for bounded mean delay; and $\tilde{\rho} < 1$ and $p_c < 1/r^3$ for bounded delay jitter, respectively. This result implies that the mean packet delay and delay jitter can go to infinity even if the system is not saturated. Based on the analysis, we define SBMD and SBDJ throughputs to be the maximum throughput that can safely guarantee bounded mean delay and delay jitter. These are arguably more sensible definitions of throughput for delay-sensitive applications. To maximize SBMD (resp. SBDJ) throughputs, the backoff factor $r$ should be carefully chosen for given $M$ and slot lengths, so that $\tau_s > \tau^*$, $\tau_{BBMD}$ (resp. $\tau_{BBDJ}$) $< \tau^*$, and $S_{BBMD}$ (resp. $S_{BBDJ}$) $= S^*$. Our results show that under many circumstances, the widely adopted binary EB where $r = 2$ yields a throughput that is far from optimum.

Together with our previous work on MPR WLANs, this paper has completed the demonstration of MPR as a powerful capacity-enhancement technique for both delay-sensitive and delay-tolerant applications. Firstly, the maximum SBMD and SBDJ throughputs are shown to scale super-linearly with MPR capability $M$ for all $M$ in ALOHA networks and for medium to large $M$ in carrier-sensing networks. That is, throughput per unit cost increases with $M$ in MPR WLANs. Secondly, the sensitivity of SBMD and SBDJ throughputs with respect to backoff factor $r$ decreases for large $M$. This implies that an MPR system is more robust against suboptimality in the selection of $r$.

In the paper, we have demonstrated the ``unsafeness'' of loading the system with an offered load higher than saturation throughput, even if the sufficient and necessary condition for bounded delay is satisfied. One interesting observation from our simulations is that the level of unsafeness differs in non-carrier-sensing and carrier-sensing networks. In one experiment, we adjust $r$ so that both ALOHA and carrier-sensing networks fall in scenario 4 defined in Fig. 3, and then load the system with an offered load higher than saturation throughput. We find that it takes quite a while for the ALOHA network to collapse, while the carrier-sensing network becomes saturated very soon. In other words, throughput higher than saturation throughput can actually be observed in ALOHA networks for a duration, but not at all in carrier-sensing network. An interesting future research direction is the study of the theory behind this phenomenon, especially the dynamic of throughput collapse in WLANs with respect to different lengths of generic time slots.

Throughout this paper, we have assumed that the arrival process is Poisson. In many cases, especially in WLAN access network, traffic does not arrive at each station according to a Poisson process. Establishing sufficient and necessary conditions for bounded mean delay and delay jitter for general traffic arrival processes is another challenging problem for future research.

APPENDIX A. PROOF OF THEOREM 1

We prove Theorem 1 by contradiction. In particular, we show that when $\tau > \tau_s$, the HOL occupancy $\rho$ exceeds 1, which violates the law of classical physics.

Before proving Theorem 1, we present the following lemma. To avoid lengthy derivation, we focus on EB-based slotted ALOHA systems with $M = 1$ in Lemma 1. The lemma, however, can be generalized to other scenarios.

**Lemma 1:** $\tilde{\rho} = \lambda \mathrm{E}[X_{ne}]$ is an increasing function of $\tau$ in EB-based slotted ALOHA systems where $T = T_{coll} = T_{succ} = T_{idle}$.

Proof: In this case,

$$\mathrm{E}[X_{ne}] = T \frac{W_0(1-p_c)+(1-rp_c)}{2(1-rp_c)(1-p_c)} = \frac{TW_0}{2(1-rp_c)} + \frac{T}{2(1-p_c)} \qquad (52)$$

It is obvious that $\mathrm{E}[X_{ne}]$ increases with $p_c$, and hence is an increasing function of $\tau$.

Case (i) $\tau < \tau^*$: In this case, it is trivially straightforward that $\tilde{\rho} = \lambda \mathrm{E}[X_{ne}]$ is an increasing function of $\tau$, as $\lambda$ also increases with $\tau$.

Case (ii) $\tau \geq \tau^*$: In this case, $\lambda$ decreases with $\tau$. That is,

$$\frac{d\lambda}{d\tau} < 0. \qquad (53)$$

Substituting (12) into (53), we get

$$N\lambda > 1. \qquad (54)$$

The derivative of $\tilde{\rho}$ with respect to $\tau$ is calculated as

$$\frac{\partial \tilde{\rho}}{\partial \tau} = \frac{\partial (\lambda E[X_{ne}])}{\partial \tau} = PL \frac{\partial}{\partial \tau} \left( \frac{\tau(W_0+r)(1-\tau)^{N-1}+\tau(1-r)}{2(1-r+r(1-\tau)^{N-1})} \right)$$
$$= PL \frac{W_0 r(1-\tau)^{2N-2}+W_0(1-\tau)^{N-2}(1-N\tau)(1-r)+2r(1-\tau)^{N-1}(1-p_c r)+(1-r)^2}{2\left(r(1-\tau)^{N-1}+1-r\right)^2} \tag{55}$$

Since $p_c < 1/r$ (which is necessary for steady state), $N\tau > 1$, and $r \geq 1$, it is easily seen that $\frac{\partial \tilde{\rho}}{\partial \tau} > 0$. Hence $\tilde{\rho}$ is an increasing function of $\tau$. When $\tau = \tau_s$, $\tilde{\rho} = 1$. Lemma 1 implies that $\tilde{\rho}$ exceeds 1 if $\tau > \tau_s$. Since $\rho \geq \tilde{\rho}$, $\rho$ also exceeds 1 when $\tau > \tau_s$, and Theorem 1 follows.

APPENDIX B. EXPRESSIONS FOR $\theta_1$, $\theta_2$, AND $\theta_3$ IN (32)

$$\theta_1 = \frac{A_1^3}{4}\left(-\frac{W_0^2}{1-r^2 p_c}+\frac{4W_0}{1-rp_c}-\frac{3}{1-p_c}\right)+\frac{A_1 A_2}{4}\left(\frac{W_0^2}{1-r^2 p_c}-\frac{6W_0}{1-rp_c}+\frac{5}{1-p_c}\right)+\frac{A_3}{2}\left(\frac{W_0}{1-rp_c}-\frac{1}{1-p_c}\right) \tag{56}$$

$$\theta_2 = \frac{A_1^3}{12}\left(\begin{array}{c}\frac{W_0^3}{2}\frac{1-r^3 p_c^2}{(1-rp_c)(1-r^2 p_c)(1-r^3 p_c)}-3W_0^2\frac{1+rp_c}{(1-rp_c)(1-r^2 p_c)}\\+\frac{11W_0}{2}\frac{1-rp_c^2}{(1-p_c)(1-rp_c)^2}-\frac{W_0^2}{2}\frac{1-r^2 p_c^2}{(1-p_c)(1-r^2 p_c)^2}-\frac{5(1+p_c)}{2(1-p_c)^2}\end{array}\right)$$
$$+\frac{A_1^2}{12}\left(\begin{array}{c}T_{coll}\left(W_0^2\frac{p_c(1+r^2-2r^2 p_c)}{(1-p_c)(1-r^2 p_c)^2}-6W_0\frac{p_c(1+r-2rp_c)}{(1-p_c)(1-rp_c)^2}+\frac{10p_c}{(1-p_c)^2}\right)\\+T_{succ}\left(W_0^2\frac{1}{1-r^2 p_c}-6W_0\frac{1}{1-rp_c}+\frac{5}{1-p_c}\right)\end{array}\right) \tag{57}$$
$$+A_1 A_2\left(\frac{W_0^2}{4}\frac{1+rp_c}{(1-rp_c)(1-r^2 p_c)}-\frac{W_0}{2}\frac{1-rp_c^2}{(1-p_c)(1-rp_c)^2}+\frac{1+p_c}{4(1-p_c)^2}\right)$$
$$+A_2\left(\frac{T_{coll} p_c}{2}\frac{W_0(1+r-2rp_c)(1-p_c)-2(1-rp_c)^2}{(1-p_c)^2(1-rp_c)^2}+\frac{T_{succ}}{2}\frac{W_0(1-p_c)-(1-rp_c)}{(1-p_c)(1-rp_c)}\right)$$

$$\theta_3 = A_1^3 \left( \begin{array}{c} \dfrac{W_0^3}{8} \dfrac{1+2rp_c+2r^2p_c+r^3p_c^2}{(1-r^3p_c)(1-r^2p_c)(1-rp_c)} - \dfrac{3W_0^3}{8} \dfrac{1+2rp_c-2rp_c^2-2r^2p_c^2-2r^3p_c^2+2r^3p_c^3+r^4p_c^4}{(1-p_c)(1-rp_c)^2(1-r^2p_c)^2} \\ + \dfrac{3W_0}{8} \dfrac{1-6rp_c^2+p_c(1+r)+rp_c^3(1+r)+r^2p_c^4}{(1-p_c)^2(1-rp_c)^3} - \dfrac{5p_c^2-4p_c+5}{8(1-p_c)^3} \end{array} \right)$$

$$+ \left( T_{coll}^3 p_c \dfrac{5p_c^2+4p_c+5}{(1-p_c)^3} + 3T_{coll}^2 T_{succ} p_c \dfrac{1+p_c}{(1-p_c)^2} + 3T_{coll} T_{succ}^2 p_c \dfrac{1}{1-p_c} + T_{succ}^3 \right)$$

$$+ 3A_1^2 \left( \begin{array}{c} \dfrac{T_{coll} W_0^2}{4} \dfrac{p_c(1-r^2p_c^2)(1-2r^2p_c+r^2)+2rp_c(1-p_c)(1-r^2p_c)}{(1-p_c)(1-rp_c)^2(1-r^2p_c)^2} \\ -T_{coll} W_0 \dfrac{p_c+rp_c+r^2p_c^2-3rp_c^2}{(1-p_c)^2(1-rp_c)^3} + \dfrac{T_{coll}}{2} \dfrac{3p_c^2-2p_c+2}{(1-p_c)^3} \\ + \dfrac{T_{succ} W_0^2}{4} \dfrac{1+rp_c}{(1-rp_c)(1-r^2p_c)} - \dfrac{T_{succ} W_0}{2} \dfrac{1-rp_c^2}{(1-p_c)(1-rp_c)^2} + \dfrac{T_{succ}}{4} \dfrac{1+p_c}{(1-p_c)^2} \end{array} \right)$$

$$+ 3A_1 \left( \begin{array}{c} \dfrac{T_{coll}^2 W_0}{2} \dfrac{rp_c-2rp_c^2-3rp_c^3+r^2p_c^2+p_c+p_c^2-3r^2p_c^3+4r^2p_c^4}{(1-p_c)^2(1-rp_c)^3} \\ +T_{succ} T_{coll} W_0 \dfrac{p_c(1+r-2rp_c)}{(1-p_c)(1-rp_c)^2} + \dfrac{T_{succ}^2 W_0}{2} \dfrac{1}{(1-rp_c)} \\ - \dfrac{T_{coll}^2}{2} \dfrac{8p_c^2-6p_c+4}{(1-p_c)^3} - T_{coll} T_{succ} \dfrac{2p_c}{(1-p_c)^2} - \dfrac{T_{succ}^2}{2} \dfrac{1}{(1-p_c)} \end{array} \right) \tag{58}$$